\documentclass[useAMS,usenatbib]{mn2e}
\usepackage{epsfig}
\usepackage{graphicx}
\usepackage{psfrag}
\newcommand{\mnras}{Monthly Notices of the Royal Astronomical Society}
\newcommand{\aj}{Astronomical Journal}
\newcommand{\apj}{Astrophysical Journal}
\newcommand{\apjl}{Astrophysical Journal Letters}
\newcommand{\aap}{Astronomy and Astrophysics}

\newcommand{\apjs}{The Astrophysical Journal Supplement Series}
\newcommand{\jcap}{Journal of Cosmology and Astroparticle Physics}
\newcommand{\physrep}{Physics Reports}
\newcommand{\prd}{Physical Review D}

\begin{document}

\title[]{
The impact of galaxy colour gradients on cosmic shear measurement
}
\author[]{L. M. Voigt$^{1}$\thanks{E-mail:lvoigt@star.ucl.ac.uk (LMV)
}, S. L. Bridle$^{1}$, A. Amara$^{2}$, M. Cropper$^{3}$,
T. D. Kitching$^{4}$, \and
R. Massey$^{4}$,  J. Rhodes$^{5,6}$, T. Schrabback$^{7}$\\
$^{1}$Department of Physics and Astronomy, University College London, Gower Street, London, WC1E 6BT, UK.\\
$^{2}$Department of Physics, ETH Z\"{u}rich, Wolfgang-Pauli-Strasse 16, CH-8093 Z\"{u}rich, Switzerland.\\
$^{3}$Mullard Space Science Laboratory, University College London, Holmbury St Mary, Dorking, Surrey RH5 6NT, UK.\\
$^{4}$University of Edinburgh, Royal Observatory, Blackford Hill, Edinburgh, EH9 3HJ, UK.\\
$^{5}$Jet Propulsion Laboratory, California Institute of Technology, 4800 Oak Grove Drive, Pasadena, CA 91109.\\
$^{6}$California Institute of Technology, 1201 E California Blvd., Pasadena, CA 91125, USA.\\
$^{7}$Kavli Institute for Particle Astrophysics and Cosmology, Stanford University, 382 Via Pueblo Mall, Stanford, CA 94305-4060, USA.\\
}

\date{Accepted . Received ; in original form }

\pagerange{\pageref{firstpage}--\pageref{lastpage}} \pubyear{2008}

\maketitle

\label{firstpage}

\begin{abstract}
Cosmic shear has been identified as the method with the most potential to constrain dark energy. To capitalise on this potential it is necessary to measure galaxy shapes with great accuracy, which in turn requires a detailed model for the image blurring by the telescope and atmosphere, the Point Spread Function (PSF). In general the PSF varies with wavelength and therefore the PSF integrated over an observing filter depends on the spectrum of the object. For a typical galaxy the spectrum varies across the galaxy image, thus the PSF depends on the position within the image. We estimate the bias on the shear due to such colour gradients by modelling galaxies using two co-centered, co-elliptical S\'{e}rsic profiles, each with a different spectrum. We estimate the effect of ignoring colour gradients and find the shear bias from a single galaxy can be very large depending on the
properties of the galaxy. We find that halving the filter width reduces the shear bias by a factor of about 5. We show that, to first order, tomographic cosmic shear two point statistics depend on the mean shear bias over the galaxy population at a given redshift. For a single broad filter, and averaging over a small galaxy catalogue from Simard et al. (2002), we find a mean shear bias which is subdominant to the predicted statistical errors for future cosmic shear surveys. However, the true mean shear bias may exceed the statistical errors, depending on how accurately the catalogue represents the observed distribution of galaxies in the cosmic shear survey. We then investigate the bias on the shear for two-filter imaging and find that the bias is reduced by at least an order of magnitude. Lastly, we find that it is possible to calibrate galaxies for which colour gradients were ignored using two-filter imaging of a fair sample of noisy galaxies, if the galaxy model is known. For a signal-to-noise of 25 the number of galaxies required in each tomographic redshift bin is of order $10^{4}$.
\end{abstract}
\begin{keywords}
galaxy shapes
\end{keywords}

\section{Introduction}
\label{sect:intro}

Cosmic shear is the weak distortion of distant galaxy images due to the weak gravitational lensing of light by intervening matter. Light rays emitted by nearby galaxies follow similar paths through the Universe and thus their images are coherently distorted. A statistical analysis of correlations in the cosmic shear signal on different scales therefore provides crucial information about the distribution of dark matter and thus the cosmological model. Several upcoming and future observational surveys plan to capitalise on the potential of cosmic shear for discovering the nature of dark energy. These include the ground-based projects the KIlo-Degree Survey (KIDS), Pan-STARRS~\footnote{\tt{http://pan-starrs.ifa.hawaii.edu}}, \emph{Subaru} Measurement of Images and Redshifts (SuMIRe)~\footnote{\tt{http://sumire.ipmu.jp/en/}} the Dark Energy Survey (DES)~\footnote{\tt{http://www.darkenergysurvey.org}} and the Large Synoptic Survey Telescope (LSST)~\footnote{\tt{http://www.lsst.org}}, and the proposed space missions \emph{Euclid}\footnote{\tt{http://sci.esa.int/euclid}} and \emph{WFIRST}\footnote{\tt{http://wfirst.gsfc.nasa.gov}}.

As surveys become more ambitious and the projected statistical uncertainties on cosmological parameters reduce, the biases associated with shear measurement must be understood and controlled with increasing accuracy. Considerable work has been done on the main systematics, including accurate shear measurement from images \citep[see][and references therein]{STEP1mnras,STEP2mnras,great08mnras,great08resultsmnras,great10handbook}, requirements on measurement and calibration of galaxy redshifts \citep{Huterer:2005ez,mahh06,bridle07,abdallaacclr08,jouvelea09,joachimib10,bernsteinh10,zhangpb10} and methods~\citep[see][and references therein]{PHAT10} and accounting for galaxy intrinsic alignments \citep{kings02,heymansh03,kings03,takada04,heymansea04,king05,bridle07,joachimi08,bernstein09,joachimis09,joachimib10,kirkbs10}.

In this paper we consider a crucial aspect of shear measurement:  deconvolution of the Point Spread Function (PSF). As light from a galaxy passes through the atmosphere and telescope optics and onto the detector it is convolved with a kernel known as the PSF. The PSF is generally not circular and we typically want to measure cosmic shear for galaxies down to the size of the PSF. The PSF must be accurately determined  either from the galaxy spectrum and a detailed model of the telescope and/or from stars in the image, which can be treated as point objects before the convolution. Several papers have put forward methods for quantifying the spatial (across the detector) and temporal variations in the shape of the PSF \citep[e.g.][]{jjb06,amaraetal10}, as well as looking at the number of noisy stars needed to accurately calibrate the PSF model \citep{paulinavrb08,paulinetal09}.

A further issue is the wavelength dependence of the PSF which becomes more important when galaxies are imaged in wide filters. The PSF is a function of wavelength and therefore the observed image consists of the galaxy image at each wavelength, convolved with the PSF at each wavelength, and then integrated over the filter response. \citet{cypriano10} investigated the first order impact of galaxy spectral energy distributions (SEDs) combined with a wavelength dependent PSF. If the spectrum of the galaxy is not known then in principle this makes it impossible to deconvolve the PSF, even if the PSF shape as a function of wavelength is perfectly known. \citet{cypriano10} have shown that the colour information required from ground-based surveys, such as the DES, for estimating galaxy photometric redshifts is enough to provide the required spectral information, if galaxies have the same spectrum at all points on their image.

In this paper we investigate the second order effect of galaxy internal colour gradients. This has been highlighted in the context of upcoming space missions \citep{cypriano10}, and Hirata \& Bernstein (priv. comm.) have shown that, if ignored, the effect can produce shear measurement biases more than an order of magnitude larger than the requirements for upcoming surveys.

Galaxies are often described by two-component models in which the bulge component is redder then the disk component. In this case there is no single PSF across the image if the PSF is a function of wavelength. Using one SED for the bulge component and another for the disk we simulate galaxy images and recover shear values using the PSF for the composite galaxy spectrum across the whole image. The resulting bias is compared to the requirement for future cosmic shear surveys to determine whether a single optical imaging filter is sufficient if galaxy color gradients are ignored. We use a catalogue of galaxies to find a reasonable range of galaxy parameters. We then repeat the calculation with two filters, allowing estimates of the bulge and disk spectra separately. Finally, we investigate the feasibility of calibrating the bias on a subset of galaxies imaged in more than one filter.

The paper is organised as follows. In Section~\ref{sect:gal_psfmod} we describe the galaxy and telescope models used in the simulations. In Section~\ref{sect:biases} we briefly outline the equations describing the distortion to galaxy images from weak gravitational shear. We then describe in detail the simulations and method used to estimate the biases on the shear and set out the requirements on the bias for future surveys. In Section~\ref{sect:single_filter} we quantify the bias on shear estimation from internal colour gradients for single filter imaging. In Section~\ref{sect:2fil} we investigate the improvement gained by imaging in two filters. We then quantify the number of galaxies needed to calibrate the biases in Section~\ref{sect:bias_calib}. Finally we discuss the implications of these results in Section~\ref{sect:discussion}.

\section{The galaxy and telescope models}
\label{sect:gal_psfmod}

Here we describe the models used to generate the galaxy and PSF images used in the simulations. We also outline the procedure for convolving the sheared multi-component galaxy images with the PSF.

\begin{figure}
\center
\epsfig{file=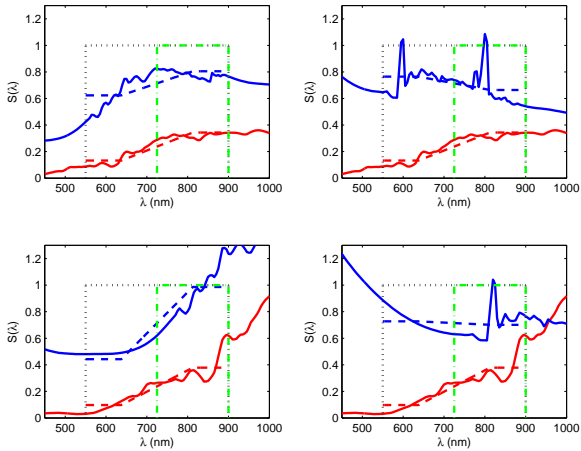,height=6cm,angle=0}
\caption{Disk (upper blue solid curves) and bulge (lower red solid curves) for a CWW-Sbc (left panels) and
CWW-Im (right panels) disk spectrum for a galaxy at redshift $z=0.6$ (upper panels) and $z=1.2$ (lower panels). The bulge spectrum is CWW-E.
The response functions $T(\lambda)$ are shown for the $F_{1}$ (black dotted) and $F_{2}$ (green dash-dotted) filters.
The dashed lines show the linear fits to the fluxes in the $F_{1}$ and $F_{2}$ filters.
}
\label{fig:spectra}
\end{figure}

\subsection{The telescope model}
\label{subsect:psfmod}

The image of a point source is broadened as a result of two dominant wavelength-dependent processes: (1) diffraction giving rise to an Airy disk with size proportional to wavelength and (2) the detector modulation transfer function (MTF) which tends to spread out higher energy photons more than lower energy photons. The PSF intensity map as a function of $\emph{\textbf{x}}$ for a galaxy with SED $S(\lambda)$ is given by
\begin{equation}
I_{\rm p}(\emph{\textbf{x}})=\frac{\int_{0}^{\infty} I_{\rm p}(\emph{\textbf{x}};\lambda) S(\lambda) T(\lambda) d\lambda}{\int_{0}^{\infty} S(\lambda) T(\lambda) d\lambda},
\label{eqn:Ipsf}
\end{equation}
where $I_{\rm p}(\emph{\textbf{x}};\lambda)$ is the normalised PSF map at each wavelength, $T(\lambda)$ is the instrumental plus filter response and \( \int_{0}^{\infty} S(\lambda) T(\lambda) d\lambda\) is the total observed flux from the galaxy.

Throughout the paper we assume a top-hat response function with $T(\lambda)=1$. For simplicity we model the PSF intensity at each wavelength, $I_{\rm p}(\emph{\textbf{x}};\lambda)$, by a Gaussian with `half-light radius' (radius enclosing half the total flux) given by
\begin{equation}
\mathrm{r}_{\rm p}(\lambda)=r_{\rm p,0} \left(\frac{\lambda}{\lambda_{0}}\right)^{0.6},
\label{eqn:fwhmp}
\end{equation}
where $\lambda_{0}=520 \rm nm$ and $r_{\rm p,0}=0.7$ pixels. The PSF parameters $\lambda_{0}$ and $r_{\rm p,0}$ are chosen so that the Full Width at Half Maximum (FWHM) of the flux-weighted PSF of the fiducial galaxy (see Section~\ref{subsect:fid_gal_mod}) is approximately 1.7 pixels. The PSF ellipticity and orientation are fixed throughout the paper at $e_{\rm p}=0.05$ and $\phi_{\rm p}=0^{\circ}$ respectively. The FWHM and ellipticity of the PSF are chosen to approximately represent the values expected in future cosmic shear surveys.

We have used an elliptical Gaussian throughout the paper to model the PSF at each wavelength. For more realistic estimates a PSF similar to the instrument in question should be used, for example an Airy disk for a space mission.

The fiducial filters we consider in this paper are a broad top-hat filter with width 550--900 nm, and a narrower filter with width 725--900 nm (hereafter $F_{1}$ and $F_{2}$ respectively). We choose $F_{1}$ to correspond to the filter currently under consideration for the \emph{Euclid} satellite.
$F_{2}$ is half the linear width of the $F_{1}$ filter and at the red end of the $F_{1}$ filter. The filter response functions are shown in Fig.~\ref{fig:spectra} along with example bulge and disk galaxy spectra, which are introduced in Section~\ref{subsect:fid_gal_mod}.

\subsection{The galaxy model}
\label{subsect:galmod}

In this paper we simulate two-component galaxies with a realistic range of profile shapes and colour gradients. The bulge and disk are modelled by S\'{e}rsic profiles \citep{sersic1968}. The S\'{e}rsic intensity at position $\emph{\textbf{x}}=(x,y)$ is given by
\begin{eqnarray}
I_{\rm g}(\emph{\textbf{x}})=I_{0} e^{-k [(\emph{\textbf{x}}-\emph{\textbf{x}}_{0})^T \emph{\textbf{C}} (\emph{\textbf{x}}-\emph{\textbf{x}}_{0})]^{\frac{1}{2n_{\rm s}}}}
\end{eqnarray}
where $\emph{\textbf{x}}_{0}$ is the centre, $I_{0}$ is the peak intensity, $n_{\rm s}$ is the S\'{e}rsic index and $\emph{\textbf{C}}$ (proportional to the inverse covariance matrix if $n_{\rm s}=0.5$) has elements
\begin{eqnarray}
C_{11}=\left(\frac{\rm cos^{2} \phi}{a^2}+\frac{\rm sin^{2} \phi}{b^2}\right)
\end{eqnarray}
\begin{eqnarray}
C_{12}=\frac{1}{2}\left(\frac{1}{a^2}-\frac{1}{b^2}\right) \rm{sin} (2\phi)
\end{eqnarray}
\begin{eqnarray}
C_{22}=\left(\frac{\rm sin^{2} \phi}{a^2}+\frac{\rm cos^{2} \phi}{b^2}\right)
\end{eqnarray}
where $\phi$ is the angle (measured anti-clockwise) between the $x$-axis and the major axis of the ellipse and the minor to major axis ratio is $b/a$.

The S\'{e}rsic index defines the profile `type', with $n_{\rm s}=0.5, 1$ and $4$ for Gaussian, exponential and de Vaucouleurs profiles respectively. If $k$ is defined as $k=1.9992n_{\rm s}-0.3271$ then for a circular profile $r_{\rm e}=a=b$ is the half-light radius (Note that for a Gaussian profile $a^{2}$ and $b^{2}$ are the 2D variances if $k=0.5$; for the circular exponential profile $h=a=b$ is known as the `scale length' when $k=1$.)

The fiducial model used in this paper for the bulge is a de Vaucouleurs profile and for the disk an exponential profile. The FWHM is related to the half-light radius (for a circular profile) via
\begin{eqnarray}
\mathrm{FWHM}=2r_{\mathrm{e}}\left(\frac{\ln 2}{
k}\right)^{n_{\rm s}}.
\end{eqnarray}
The total flux (integrated to infinity) emitted by a galaxy described by a S\'{e}rsic profile with index $n_{\rm s}$ is given by
\begin{equation}
F=2\pi n_{\rm s} k^{-2n_{\rm s}}r_{\rm e}^{2} \Gamma(2n_{\rm s}) I_{0}
\label{eqn:flux}
\end{equation}
where $\Gamma$ is the gamma function.

\begin{figure*}
\center
\epsfig{file=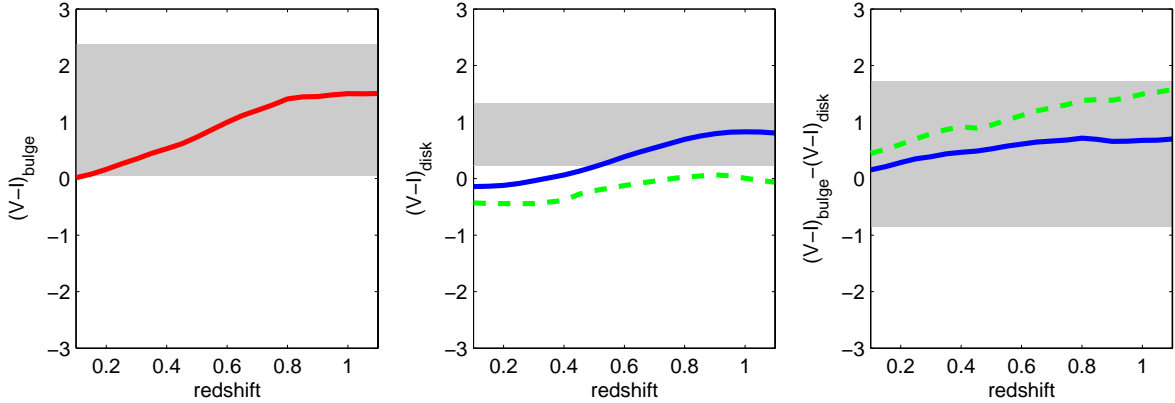,height=5.5cm,angle=0}
\caption{Bulge colour (left), disk colour (middle) and bulge minus disk colour (right) as a function of redshift for a CWW-Sbc (blue solid) and CWW-Im (green dashed) disk spectrum. The bulge spectrum is CWW-E (red solid). Fluxes are measured in the $V$ and $I$ bands. The shaded regions show the central 68 per cent of the colour distribution measured from galaxies in the catalogue (see Section~\ref{subsect:gal_cat}).
}
\label{fig:catalogue_Im_Sbc}
\end{figure*}

\subsection{The galaxy catalogue}
\label{subsect:gal_cat}

The results will depend on the exact assumptions made about the galaxy properties and therefore we use a galaxy catalogue to estimate the average quantities required. The catalogue we use is from a study by \citet{simard02} of \emph{Hubble Space Telescope}
(\emph{HST}) WFPC2 F606W (470--720 nm; `$V$-band') and F814W (708--959 nm; `$I$-band') observations of the `Groth Strip'. The catalogue lists galaxy photometric structural parameters obtained from a two-dimensional bulge plus disk decomposition of the galaxy surface brightness profiles. The bulge profile is modelled as a de Vaucouleurs and the disk as an exponential.

\citet{simard02} produce two sets of catalogues, one in which fits are performed separately in each filter, and a second catalogue in which the fits are performed simultaneously in each bandpass. We use the second catalogue, which is from the DEEP1 (Deep Extragalactic Evolutionary Probe) Keck LRIS spectroscopic survey of the Groth Strip \citep{vogtetal05}. The sample is magnitude-limited and contains several objects which were prioritized for selection \citep[see][]{weineretal05}. The catalogue contains 632 galaxies in the magnitude (Vega) range $(V+I)/2 < 24$ and the median redshift is 0.65.

The catalogue is well-suited to our study because the broad $V$ and $I$ bands used in the \emph{HST} observations cover a similar wavelength range to the broad $F_{1}$ filter.The catalogue provides bulge and disk scale radii (fixed at the same value in each filter) and the magnitudes of the bulge and the disk in each filter.

We compute linear approximations to the bulge and disk spectra, $S(\lambda)$, across the $V$ and $I$ bands from the fluxes in each filter using the $AB$ magnitudes given in the catalogue\footnote{$m=-2.5 \mathrm{log_{10}}(F)-48.6$, where $m$ is the $AB$ magnitude and $F$ is the flux.}. The bulge SED between the midpoints of the $V$ and $I$-bands (595--833.5 nm) is given by $S(\lambda)=m_{\rm b}\lambda+c_{\rm b}$, where $m_{\rm b}$ and $c_{\rm b}$ are the gradient and $y$-intercept of the spectrum respectively, and similarly for the disk. The spectrum is assumed to be flat between 470--595 nm and 833.5--959 nm, such that $S(\lambda)=m_{\rm b}\lambda_{V_{\rm mid}}+c_{\rm b}$ in the first half of the $V$-band and $S(\lambda)=m_{\rm b}\lambda_{I_{\rm mid}}+c_{\rm b}$ in the second half of the $I$-band, where $\lambda_{V_{\rm mid}}=595$nm and $\lambda_{I_{\rm mid}}=833.5$nm. The gradient and intercept of the bulge and disk spectra are computed so that the total bulge and disk fluxes in the $V$ and $I$ bands are equal to the values in the catalogue. If $S(\lambda)$ goes negative in either band then the spectrum is re-set to be flat with a different value on either side of 720 nm, with the constraint that the flux between 470--720 nm is equal to the flux in the $V$-band, and the flux between 720--959 nm is equal to the flux in the $I$-band minus the flux between 708--720 nm calculated from the flat spectrum between 470--720 nm.

The exact distribution of galaxies in a particular survey will depend on the parameters of the survey (e.g. the median redshift), and this must be taken into account when using the results obtained from the catalogue described here to make predictions for future cosmic shear surveys. We investigate the sensitivity of the results to the galaxy redshift and colour distributions in the catalogue in Section~\ref{subsect:averages}.

We also note that the photometric and structural parameters obtained in the \citet{simard02} study will not precisely represent the underlying distribution of galaxy colours and shapes in the DEEP1 survey. This is because (i) the study fits a simple two-component model to the galaxy images, when in fact galaxy morphologies may be complex and (ii) galaxy images are noisy which may bias the fitted parameters \citep[see][]{haussleretal07}.

\subsection{The fiducial galaxy model}
\label{subsect:fid_gal_mod}

We also consider a fiducial galaxy model which is the `average' galaxy from the catalogue. The fiducial galaxy parameters are $n_{\rm s,b}=4$, $n_{\rm s,d}=1$, $r_{\rm e,b}/r_{\rm e,d}=1.1$, $B/T=0.25$ and $z=0.9$. The bulge and disk ellipticities are set equal to $e_{\rm b}=e_{\rm d}=e_{\rm g}=0.2$. The fiducial values chosen for the ratio of the bulge-to-disk half light radii and the bulge-to-total flux ratio are the mean values from the catalogue. The central 68\% of the $r_{\rm e,b}/r_{\rm e,d}$ distribution is approximately $1.1^{+0.5}_{-0.8}$. The fiducial redshift ($z=0.9$) is the median value for \emph{Euclid}.

We use a CWW-E spectrum for the bulge throughout the paper, and either a CWW-Sbc or CWW-Im spectrum for the disk \citep[see][]{colemanetal1980}. We use the CWW-Sbc spectrum for the fiducial disk spectrum. The spectra observed in the wavelength range 500--1000 nm are shown in Fig.~\ref{fig:spectra} for two different source redshifts. We show in Fig.~\ref{fig:catalogue_Im_Sbc} that these spectra provide a good representation of the distribution of bulge and disk colours found in the catalogue, with the CWW-Sbc spectrum representing the disk $V-I$ colours more closely than the CWW-Im spectrum.

The bulge and disk half-light radii are rescaled (keeping the ratio of the bulge half-light radius to disk half-light radius constant at the fiducial value) throughout the paper so that the ratio of the PSF-convolved galaxy FWHM to PSF FWHM, $r_{\rm gp}$, is fixed at 1.4 for a circular PSF and galaxy.

\subsection{Convolution}
\label{subsect:convolution}

The PSF-convolved galaxy intensity at \emph{\textbf{x}} is given by
\begin{equation}
I_{\rm g*p}(\emph{\textbf{x}})=\int_{0}^{\infty} I_{\rm g}(\emph{\textbf{x}};\lambda) * I_{\rm p}(\emph{\textbf{x}};\lambda) T(\lambda) d\lambda
\end{equation}
where $I_{\rm g}(\emph{\textbf{x}};\lambda)$ is the galaxy intensity per unit wavelength.
For multiple component galaxies with components $i$ this can be written as
\begin{eqnarray}
I_{\rm g*p}(\emph{\textbf{x}})
&=& \sum_{i} I_{\rm g,i}(\emph{\textbf{x}}) * I_{\rm p,i}(\emph{\textbf{x}}) \\
I_{\rm p,i}(\emph{\textbf{x}})
&=&
\frac{\int_{0}^{\infty} S_{\rm i}(\lambda) T(\lambda) I_{\rm p}(\emph{\textbf{x}};\lambda) d\lambda}{\int_{0}^{\infty} S_{\rm i}(\lambda) T(\lambda) d\lambda}
\end{eqnarray}
where $I_{\rm g,i}(\emph{\textbf{x}})$ is the galaxy intensity integrated over all wavelengths and $S_{i}(\lambda)T(\lambda)d\lambda$ is the
total observed flux from the $i$th component per unit wavelength.

\section{Quantifying the bias on shear measurements}
\label{sect:biases}

We quantify the bias on cosmic shear measurements from using an incorrect PSF model for a given set of galaxy parameters.
We use the `ring-test' method described in \citet{nakajimab07} \citep[see also][]{voigt&bridle10}.
We first briefly outline the equations describing the distortions to galaxy images from gravitational shear before describing the simulations in more detail.

\subsection{Gravitational Shear}
\label{subsect:grav_shear}

The two-component gravitational shear of a galaxy, given by
\begin{equation}
\gamma=\gamma_{1}+i\gamma_{2},
\end{equation}
\label{eqn:gamma}
is related to the projected dimensionless gravitational potential $\psi$ of a thin lens between the source and the observer such that
\begin{equation}
\gamma_{1}=\frac{1}{2}
\left(\psi_{11}-\psi_{22}\right),
\gamma_{2}=\psi_{12}=\psi_{21},
\end{equation}
where \(\psi_{ij}=\partial^{2}\psi/\partial \theta_{i} \partial \theta_{j}\) and $\btheta$ is the observed position of the source \citep[see, for e.g.,][]{bartelmann&schneider01}. In the weak lensing regime the reduced shear, $g$, is approximately equal to the shear $\gamma$ where
\begin{equation}
g=\gamma/(1-\kappa),
\end{equation}
and the convergence $\kappa$ is proportional to the projected surface mass density of the lens. A pre-shear elliptical isophote galaxy image with minor to major axis ratio $b/a$ and orientation of the major axis anticlockwise from the positive $x$-axis, $\phi$, has a complex ellipticity
\begin{equation}
{e}^{\mathrm{s}}
=\frac{a-b}{a+b}
e^{2i\phi},
\end{equation}
\label{eqn:defe}
and is sheared to an ellipse with complex observed ellipticity $e^{\mathrm{o}}$ given by
\begin{equation}
e^{\mathrm{o}}=\frac{e^{\mathrm{s}}+g}{1+g^{*}e^{\mathrm{s}}}
\label{eqn:eobs}
\end{equation}
\citep{seitz97} if $\kappa<1$. We will assume $\kappa<<1$ throughout this paper.

\subsection{The Simulations}
\label{subsect:simulations}

We simulate two-component galaxies at different orientations with co-centered, co-elliptical profiles. Each galaxy is sheared by the same amount $g$. The estimated shear for each galaxy is the measured observed ellipticity. We use a ring-test to quantify the bias on the estimated shear for a given galaxy model and PSF estimate in the limit of an infinite number of orientations of the pre-sheared galaxy. For a perfect shear measurement method the average observed ellipticity is then equal to the true input shear. In practice we can reduce the number of galaxies needed in the ring-test by simulating pairs of galaxies separated by 90 degrees \citep[as in][]{STEP2}. It can be shown from Eqn.~\ref{eqn:eobs} that even for a perfect shear measurement method we need at least 3 pairs of galaxies in the ring-test to reduce the bias on the estimated shear to a negligible level.
If the shear measurement method is imperfect (e.g. because the PSF model is incorrect) then we may need more than 3 pairs to obtain a shear value that has converged to the value we would measure for an infinite number of galaxy orientations. We find, however, that 3 pairs of galaxies is sufficient to reach convergence. We check that doubling the number of pairs of galaxies does not change the results.

PSF-convolved galaxy images are produced on postage stamps $15^{2}$ pixels in size using the following procedure. The PSF and galaxy models are described in Sections~\ref{subsect:psfmod} and~\ref{subsect:galmod} respectively. The complex ellipticities of the post-shear bulge and disk are computed using Eqn.~\ref{eqn:eobs}. The size of each component ($r_{\rm e,b}^{2}$ and $r_{\rm e,d}^{2}$) is rescaled by a factor $(1-|g|^{2})^{-1}$ after shearing. The PSF convolutions are performed on a high resolution image which is $17^2$ observational pixels in area and has each observational pixel divided into $7^2$ subpixels. We find that increasing the number of subpixels used in the convolution beyond $7^2$ has a negligible impact on the results. We find that it is necessary to create the galaxy image at an even higher resolution before binning up to make the galaxy image used in the convolution. We find it is sufficient to do this only on the central $5^2$ subpixels of the convolution image, but that each of these subpixels must be divided into $25^2$ subsubpixels. The intensity in each subpixel is found by calculating the intensity at the centre of each subsubpixel and summing all subsubpixels within each pixel. The intensity in the remaining subpixels of the image is assumed to be equal to the value at the centre
of that subpixel.
Integration of the intensity within
each of
the central $5^2$ subpixels of the image is necessary for highly peaked profiles, such as the de Vaucouleurs profile used to represent the bulge surface brightness distribution. The convolution with the PSF is performed using FFTs using sufficient padding such that the result of the convolution is identical to that from performing the convolution in real space. The high resolution grid is larger than the postage stamp to allow for light scattered back into the postage stamp due to convolution with the PSF. We check that increasing the resolution parameters does not change the results.

The high resolution PSF-convolved bulge and disk images are then binned up by a factor of seven and cut down to produce an image on a postage stamp $15^{2}$ pixels in size. The observed galaxy image is the linear sum of the bulge and disk PSF-convolved images. The bulge and disk are co-centric and co-elliptical and the peak of the intensity profile is within the central pixel of the image grid. The fiducial centre used is $\{x_{0},y_{0}\}=\{0.1,0.3\}$ pixels relative to the centre of the postage stamp for the galaxy aligned along the $x$-axis. (In the ring-test the galaxy is rotated around the centre of the grid). The PSF FWHM is 1.7 pixels and the FWHM of the PSF-convolved galaxy image is 1.4 times the PSF FWHM.

The best-fit parameters of the bulge and disk components are computed on the image (low resolution) grid for each angle in the ring test by minimising the $\chi^{2}$ between the true PSF-convolved galaxy image and the PSF-convolved galaxy image using the estimated PSF model. This process is repeated for two different input shears: $\{\gamma^{\rm t}_{1},\gamma^{\rm t}_{2}\}=\{0,0\}$ and $\{\gamma^{\rm t}_{1},\gamma^{\rm t}_{2}\}=\{0.01,0.02\}$, where the superscript `t' refers to the true value of the shear. For each input shear we obtain a shear estimate, $\hat{\gamma}$. We quantify the bias on the shear estimator in terms of multiplicative and additive errors, $m_i$  and $c_i$ respectively, following~\citet{STEP1mnras}, such that
\begin{equation}
\hat{\gamma}_i=(1+m_i) \gamma^{\rm t}_i+c_i
\end{equation}
where the subscript $i$ refers to the two shear components and we assume there is no cross contamination of e.g. $\hat{\gamma_1}$ depending on the value of $\gamma^{\rm t}_2$ or vica versa.

\subsection{Survey requirements}
\label{sect:req}

We use the requirements on the multiplicative and additive shear biases for cosmic shear two point statistics computed in \citet{amarar08} \citep[see also][]{kitchingetal09}. These are set so that the systematic error is equal to the statistical error,   where the statistical uncertainties depend on the survey area, depth and galaxy number density. For a medium-deep survey of 20,000 square degrees (hereafter `\emph{Euclid}-like survey') the upper limits on the total contribution to the multiplicative and additive biases are $10^{-3}$ and $3\times10^{-4}$ respectively. Biases a factor of 1--2, 2--5 and 5--20 below the \emph{Euclid}-like survey requirements are shown in the figures by light, medium and dark grey shaded regions respectively.

The requirements on the shear bias for a particular cosmic shear survey as computed in \citet{amarar08} must encompass all systematics involved in the survey (e.g. shear measurement, charge transfer inefficiency or CTE, intrinsic alignments etc.), and thus to be safe for a \emph{Euclid}-like survey we require the bias from the colour dependence of the PSF to be of order a factor of 10 below the upper limits on the multiplicative and additive biases stated above.

\section{Single filter imaging}
\label{sect:single_filter}

We calculate shear measurement biases for imaging in a single filter, first considering the dependence of the bias on filter width for  some example galaxies, and then calculating the mean bias for single broad-filter imaging by averaging over the catalogue.

\subsection{Dependence on filter width and galaxy properties}
\label{subsect:fil_width}

We first compute the multiplicative and additive biases on the shear for a two-component galaxy
imaged in a single visible filter of increasing width with central wavelength 725nm.
For single filter imaging we have no resolved colour information;
we assume however that the galaxy spectrum from the composite bulge plus disk image is known perfectly from unresolved observations in multiple filters.
Such observations are already required for estimating galaxy photometric redshifts.
The PSF model is described in Section~\ref{subsect:psfmod}.
We use the PSF model obtained using the SED from the composite galaxy spectrum
(equal to the flux-weighted sum of the true bulge and disk PSF models) for both the bulge and disk PSFs.
The galaxy model is described in Section~\ref{subsect:galmod}.
The bulge-to-total flux ratio is kept constant as the filter width increases. Variations in the shear bias values are
thus dominated by the increase in filter width, and not by a change in the galaxy shape.

The biases are computed using the ring-test method described in Section~\ref{subsect:simulations}.
We assume the galaxy model is known.
There are seven free parameters in the fits to the single filter galaxy images:
$x_{0}, y_{0}, e_{\rm g}, \phi_{\rm g}, r_{e, \rm b}, r_{e, \rm d}$ and $I_{0, \rm b}$,
where $I_{0, \rm b}$ is the peak intensity of the bulge profile.
Note that the peak intensity of the disk profile is known from $r_{e, \rm b}, r_{e, \rm d}, I_{0, \rm b}$ and the total galaxy flux.
We assume the total galaxy flux would be obtained from other observations such as the unresolved observations used to obtain the composite spectrum.

\begin{figure}
\center
\epsfig{file=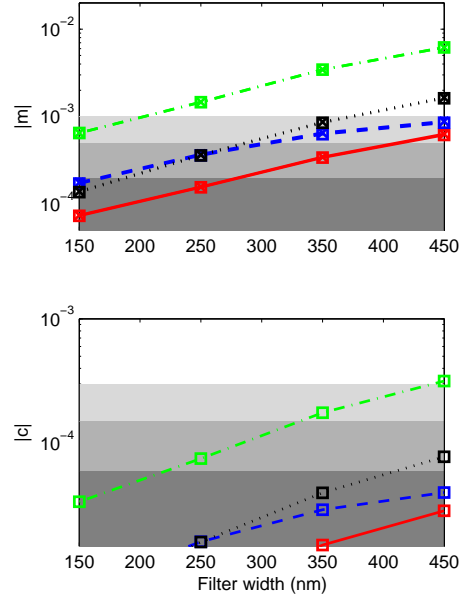,height=8cm,angle=0}
\caption{Absolute value of the multiplicative (top) and additive (bottom) shear bias
as a function of filter width for imaging in a single filter.
The filter transmission function is a top-hat and the central wavelength of the filter is 725 nm. Open squares (crosses) show $m_{1}$,$c_{1}$ ($m_{2}$,$c_{2}$).
The fiducial galaxy parameters are $z=0.9, r_{\rm e,b}/r_{\rm e,d}=1.1, B/T=0.25$, a CWW-Sbc disk spectrum and a CWW-E bulge spectrum (red solid).
Results are also shown for $z=1.4$ (blue dashed), $r_{\rm e,b}/r_{\rm e,d}=0.4$ (green dot-dashed) and a CWW-Im disk spectrum (black dotted),
with all other parameters kept at the fiducial values above.
Note that $c_{2}$ is zero because the PSF is aligned along the $x$-axis.
The shaded regions from light to dark indicate where $m_{i}$ is a factor of 1--2, 2--5 and 5--20 below the \emph{Euclid} requirement on the bias.
}
\label{fig:bias_filterwidth}
\end{figure}

\begin{figure}
\center
\epsfig{file=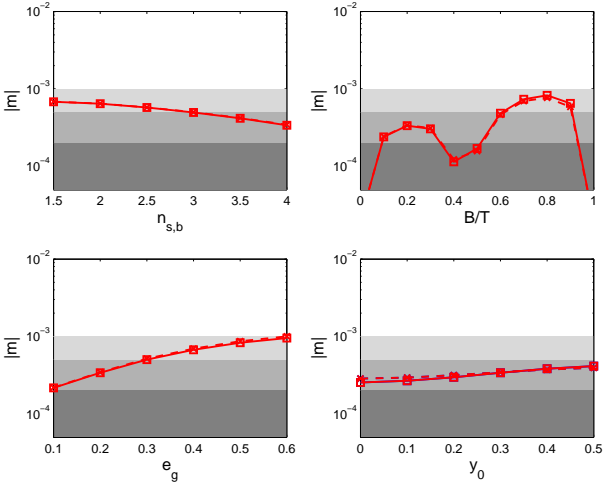,height=6.5cm,angle=0}
\caption{
Absolute value of the multiplicative shear bias
for imaging in the $F_{1}$ filter
as a function of the galaxy parameter (clockwise from top left): bulge S\'{e}rsic index, bulge-to-total flux ratio, galaxy ellipticity,
position of peak intensity relative to the centre of the postage stamp (for the galaxy aligned along the $x$-axis; see Section~\ref{subsect:simulations}; $x_{0}=0.1$).
All other galaxy parameters are fixed at the fiducial values (see Section~\ref{subsect:fid_gal_mod}).
The true bulge and disk spectra are CWW-E and CWW-Sbc respectively.
The estimated PSF model for the bulge and the disk components is the model from the composite galaxy spectrum.
Solid open squares (dashed crosses) show $m_{1}$ ($m_{2}$).
Shaded regions as in Fig.~\ref{fig:bias_filterwidth}.
}
\label{fig:galpar}
\end{figure}

Results are shown in Fig.~\ref{fig:bias_filterwidth} for the fiducial galaxy (red solid lines),
as well as for a CWW-Im disk spectrum (black dotted lines) and for a different value of the galaxy redshift
(blue dashed lines) and $r_{\rm e,b}/r_{\rm e,d}$ ratio (green dot-dashed lines).
The shear biases decrease as the filter width decreases because
the bulge and disk spectra are in general more similar for smaller widths
(and identical for vanishingly small widths). For
surveys using wide-band imaging with a filter width of 350 nm
(the $F_1$ filter)
the multiplicative biases for the fiducial galaxy considered here are a factor of
3 below the \emph{Euclid}-like survey requirement. The additive bias is
a factor of 20 below the requirement. The size of the bias is strongly affected by the ratio of the bulge to disk half-light radius,
as well as the internal colour gradient, with the multiplicative bias increasing to a factor of $\sim$4
above the requirement and additive bias a factor of $\sim$2 below the requirement for the single departure from fiducial of $r_{\rm e,b}/r_{\rm e,d}=0.4$.
We note that the distribution of bulge to disk half-light radii in the galaxy catalogue described in Section~\ref{subsect:gal_cat} has mean $r_{\rm e,b}/r_{\rm e,d}=1.1$,
and the central 68\% of the distribution is in the range $r_{\rm e,b}/r_{\rm e,d}=0.3-1.6$.

If we simultaneously make the above multiple departures from the fiducial galaxy the biases are compounded,
resulting in a multiplicative bias
30--50 times
larger than shown by the red solid curve, depending on the
filter width (e.g. for a filter width of 350 nm, $m\sim10^{-2}$ for $z=1.4$, $r_{\rm e,b}/r_{\rm e,d}=0.4$,
a CWW-Im disk spectrum
and a CWW-E bulge spectrum).
For our co-elliptical simulations the multiplicative biases dominate the additive biases and thus we consider only the multiplicative biases in the rest of the paper.
We note that for the fiducial galaxy parameters the multiplicative bias increases (decreases) at each filter width by a factor of about 1.5 when the
PSF ellipticity is doubled (halved) from the fiducial value of $e_{\rm p}=0.05$.
The additive bias increases (decreases) at each filter width by a factor of about 3 when the PSF ellipticity is doubled (halved) from the fiducial value.

We now show how different galaxy parameters affect the bias. In Fig.~\ref{fig:galpar} we plot the multiplicative shear biases as a function of bulge S\'{e}rsic index, bulge-to-total flux ratio, galaxy ellipticity and position of peak intensity within the central pixel of the postage stamp.  We use the same fiducial values as in Fig.~\ref{fig:bias_filterwidth} with a filter width of 350nm ($F_{1}$ filter). The disk profile is an exponential.
We see that changing these parameters within the ranges shown does not increase the bias above the \emph{Euclid}-like survey requirement.
The bias is little affected by the position of the galaxy.
However, we see that to determine whether single
broad
filter imaging will give biased shear measurements we need to know the true distribution of galaxy shapes and colours in the Universe.

\subsection{Averages over the galaxy catalogue}
\label{subsect:averages}

As shown in Section~\ref{subsect:fil_width}, the biases on the shear depend on the bulge and disk spectra, the galaxy redshift, the ratio of the bulge-to-disk half-light radii,
the bulge-to-total flux ratio, the galaxy ellipticity and the bulge S\'{e}rsic index.
In this section we compute the biases for single
broad-filter
imaging with the $F_{1}$ filter for a realistic range of galaxy parameters obtained from a catalogue, described in Section~\ref{subsect:gal_cat}. The catalogue represents an observed range of galaxy colour gradients and bulge-to-disk half-light radii. The bulge S\'{e}rsic index is $n_{\rm s, b}=4$ and we fix the galaxy ellipticity at $e_{\rm g}=0.2$ for all galaxies in the catalogue.
We see from Fig.~\ref{fig:galpar} that by fixing the bulge S\'{e}rsic index and galaxy ellipticity we may over- or under-estimate the mean bias by a factor of at most 3, depending on the true distribution of these parameters in the Universe.

For each galaxy in the catalogue we
simulate
separate
`true' bulge and disk PSF images in the $F_{1}$ filter by substituting the linear spectra obtained from the
bulge and disk magnitudes (see Section~\ref{subsect:gal_cat}) into Eqn~\ref{eqn:Ipsf}. Bulge and disk galaxy images are computed and convolved with the PSF images as described in Section~\ref{subsect:simulations}.
The bulge and disk shape parameters $r_{\rm e,b}/r_{\rm e,d}$ and $B/T$ are taken from the catalogue.
The bulge and disk half-light radii are scaled so that $r_{\rm gp}=1.4$.
In the fits to the `true' PSF-convolved galaxy images we use the PSF image from the composite galaxy spectrum to estimate
both the bulge and disk PSF images.
We measure the bias on the shear using a ring-test (as described in Section~\ref{subsect:simulations}) for each galaxy in the catalogue.
We assume the galaxy model is known and fit for 7 free parameters, described in Section~\ref{subsect:fil_width}.

\begin{figure}
\center
\epsfig{file=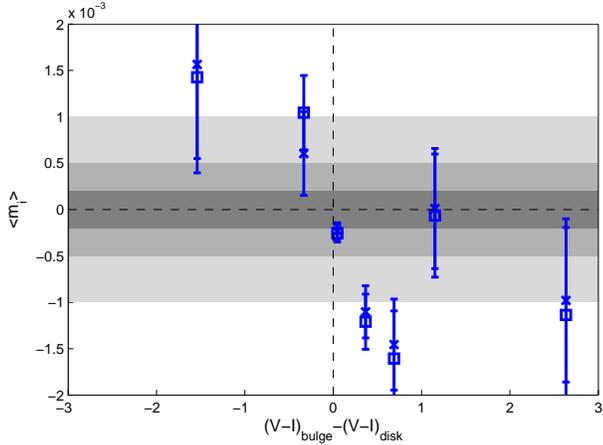,height=6cm,angle=0}
\caption{Mean multiplicative bias ($m_{1}$ open squares; $m_{2}$ crosses) as a function of
$V-I$
colour gradient using galaxies from the catalogue (see Section~\ref{subsect:gal_cat}).
Each colour gradient bin contains about 90 galaxies.
The
galaxy
images
are
simulated
in a single broad filter ($F_{1}$) with a top-hat transmission function.
The estimated PSF model for the bulge and the disk components is the model from the composite galaxy spectrum.
The error bars show the error on the mean bias in each bin (i.e. $\sigma_{m_i}/\sqrt{n_{\rm g}}$, where $\sigma_{m_i}$ is the standard deviation of the
$m_i$ distribution and $n_{\rm g}$ the number of galaxies in the bin).
Vertical (horizontal) black dashed lines show where the colour gradient ($\langle m_{1} \rangle$) is zero. Shaded regions as in Fig.~\ref{fig:bias_filterwidth}.
}
\label{fig:bias_colgrad}
\end{figure}

In Fig.~\ref{fig:bias_colgrad} we plot the multiplicative biases as a function of the galaxy internal $V-I$ colour gradient
(i.e. bulge $V-I$ colour minus disk $V-I$ colour).
The bias scales approximately linearly with the colour gradient; the shear is over-estimated for negative colour gradients (blue-cored galaxies) and under-estimated for
positive colour gradients (red-cored galaxies).
The mean bias $\langle m_{1} \rangle$ over the 632 galaxies in the catalogue is $\sim3\times10^{-4}$, a factor of about 3 below a \emph{Euclid}-like survey requirement,
and similar to the value obtained from the fiducial galaxy which was chosen to have average parameters from the catalogue.
The bias does not change monotonically with colour gradient because the ratio of bulge to disk size is also changing within the catalogue.

\begin{figure}
\center
\epsfig{file=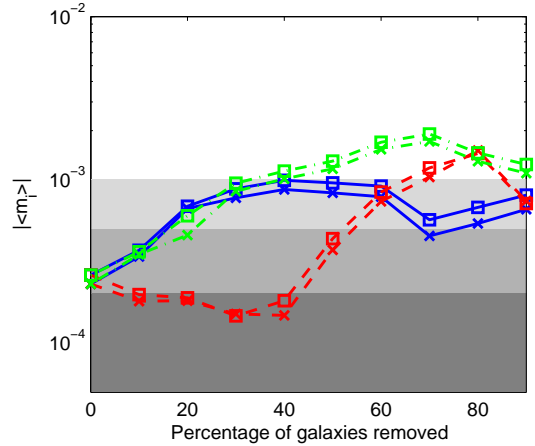,height=6cm,angle=0}
\caption{Absolute value of the mean multiplicative bias ($m_{1}$ open squares; $m_{2}$ crosses) as a function of the percentage of galaxies removed from the catalogue,
starting with galaxies with the most negative colour gradient (blue solid), most positive colour gradient (red dashed) and most negative bulge colour (green dot-dashed). Shaded regions as in Fig.~\ref{fig:bias_filterwidth}.
}
\label{fig:bias_perc_removed}
\end{figure}

We
now
investigate the sensitivity of the mean bias to the colour
distribution of galaxies in the catalogue.
In Fig.~\ref{fig:bias_perc_removed} we plot the mean bias as a function of the percentage of galaxies removed from the catalogue.
We investigate the sensitivity to the colour gradient distribution by progressively removing galaxies with the most negative colour gradient, and similarly for the most positive colour gradient (red dashed).
The mean bias is equal to the mean bias over the whole catalogue for 0\% removal. If we remove galaxies with the most negative colour gradients (blue-cored galaxies,
shown by the blue solid line)
then the mean bias increases relative to the mean bias over the complete catalogue by a factor of at most 3.
If we progressively remove galaxies with positive colour gradients (red-cored galaxies
shown by the red dashed line),
then the mean bias initially decreases before increasing to the \emph{Euclid}-like requirement for removal of 60\% or more of the catalogue.
We note that removing galaxies with the most negative (or most positive) colour gradients
increases the mean bias because the
distribution of colour gradients in the catalogue is relatively symmetric about zero colour gradient,
and the bias scales approximately linearly with colour gradient.

We also consider the sensitivity of the catalogue to the galaxy redshift distribution. For a \emph{Euclid}-like survey with median redshift $z=0.9$ the 4000{\AA} Balmer break is within the $F_{1}$ filter, thus we expect to observe a significant fraction of red bulges. We plot the mean bias as a function of the percentage of the `blue-est' (most negative $V-I$ colour) bulges removed from the catalogue. The mean bias increases to a factor of $\sim$2 above the \emph{Euclid}-like requirement for removal of 60\% of the blue-est galaxies from the catalogue. Thus if red bulges are under-represented by the catalogue relative to the fraction expected in a
\emph{Euclid}-like survey the mean bias for single
broad-filter
imaging may
be close to the statistical equals systematic error requirement of $1\times10^{-3}$.

As discussed in Section~\ref{sect:req}, the mean bias needs to be
a factor of 10 or more
below the statistical equals systematic error requirement
to be safe for a \emph{Euclid}-like survey,
and thus we consider options for reducing the bias below that for single broad-filter imaging.
As shown in Section~\ref{subsect:fil_width}, the mean bias from single broad-filter imaging may be reduced by imaging in a narrower filter.
Alternatively, it may be possible to reduce the mean bias from single broad-filter imaging to the required level by imaging in an additional filter,
or by calibrating the bias
on a sub-set of galaxies.

\begin{figure*}
\center
\epsfig{file=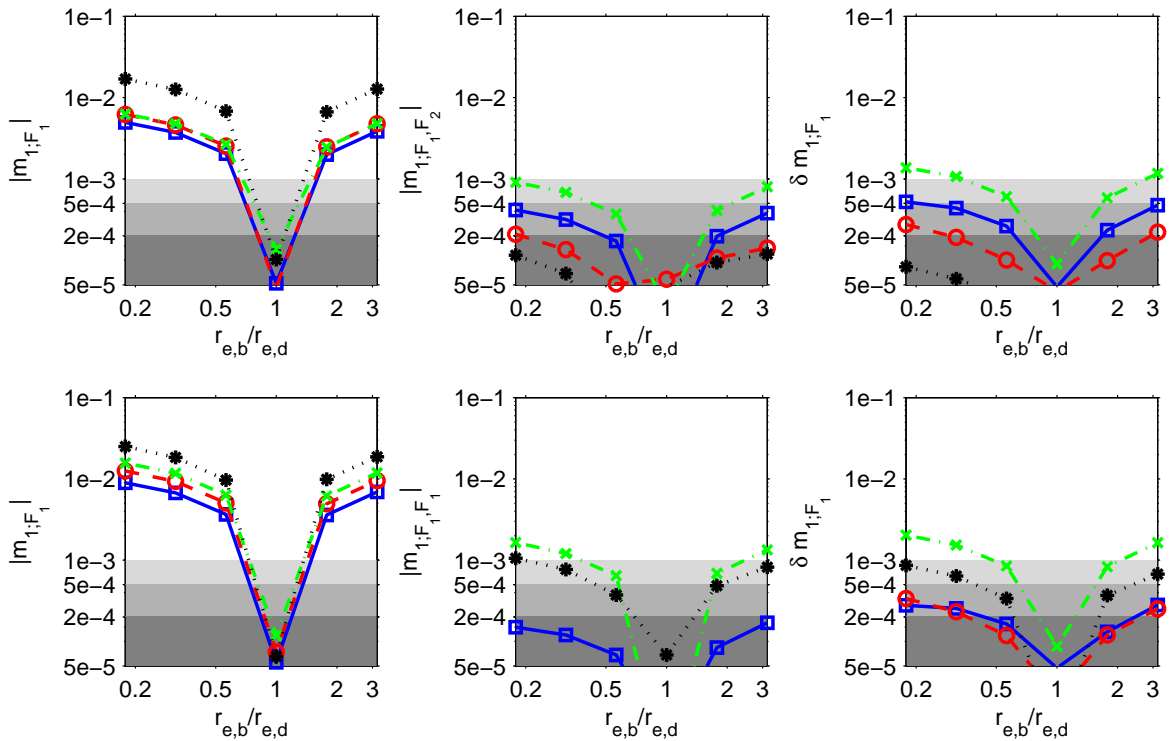,height=10cm,angle=0}
\caption{Absolute value of the multiplicative shear bias, $|m_{1}|$, as a function of the ratio of the bulge-to-disk half-light radii for CWW-Sbc (upper plots) and CWW-Im (lower plots) disk spectra for a galaxy at redshift 0.4 (blue solid squares), 0.8 (red dashed circles), 1.2 (green dot-dashed crosses) and 1.6 (black dotted stars). Results are shown for imaging in one broad filter ($F_{1}$; left-hand plots) and in two filters ($F_{1}$, $F_{2}$; middle plots). The right-hand plots show the magnitude of the difference between the multiplicative bias obtained by imaging in one filter ($F_{1}$) and the estimated bias from imaging in this filter using the bulge and disk shape parameters and spectra obtained from the two filter fits.
The
bulge-to-total
flux ratio is $B/T=0.25$ and the galaxy ellipticity is $e_{\rm g}=0.2$.
The bias values for two-filter imaging and a CWW-Im spectrum (lower middle plot) lie below the lower limit on the $y$-axis for $z=0.8$.
Shaded regions as in Fig.~\ref{fig:bias_filterwidth}.
}
\label{fig:bias_calibration_2filter}
\end{figure*}

\section{Two filter imaging}
\label{sect:2fil}

In this section we investigate how far the biases on the shear may be reduced using two filter imaging. We repeat the ring-test calculation in Section~\ref{subsect:fil_width} using a narrower filter, $F_{2}$ (725--900nm), in addition to the broad $F_{1}$ filter. The narrower filter covers the second half of the wavelength range of the wider filter (see Fig.~\ref{fig:spectra}).

The `true' PSF-convolved galaxy images are generated as described in Section~\ref{subsect:fil_width} using a CWW-E spectrum for the bulge and a CWW-Sbc or CWW-Im for the disk.
For two filter imaging we estimate separate spectra for the bulge and disk components by fitting to the observed two-filter image
and using information from the composite galaxy spectrum.
We use different PSF models for each of the bulge and disk components ($I_{\rm p,b}^{\rm est}$ and $I_{\rm p,d}^{\rm est}$ respectively),
according to their estimated spectra.

We first create linear spectra in which the bulge spectrum is assumed to be flat between
$\lambda_{F1, 0}$ and $\lambda_{F1, 0.25}$ and also from $\lambda_{F1, 0.75}$ and $\lambda_{F1, 1}$, where $\lambda_{F1, f}$
is the wavelength a linear fraction $f$ across the $F_1$ filter, for example $\lambda_{F1, 0.25}=637.5$
nm.
Therefore $S_b(\lambda)=m_{\rm b}\lambda_{F1, 0.25}+c_{\rm b}$ in the first quarter of the $F_{1}$ filter and
$S_b(\lambda)=m_{\rm b}\lambda_{F1, 0.75}+c_{\rm b}$ in the last quarter of the $F_{1}$ filter (i.e. the second half of $F_{2}$), and similarly for the disk.
The gradient and intercept of the bulge and disk spectra are
re-computed
at each iteration in the fit
so that the
estimated
bulge and disk fluxes in $F_{1}$ and $F_{2}$ are equal to the integrated fluxes from the
linear
spectra.
If $S_i(\lambda)$ from the linear fits to the fluxes
goes
negative in either band then the spectrum is re-set to be flat
with a different value
on either side of the midpoint of $F_{1}$.
Example linear fits to the bulge and disk fluxes are shown in Fig.~\ref{fig:spectra} for redshifts 0.6 and 1.2. In both cases the bulge-to-total flux ratio is 0.25.

Using just the linear spectra fitted to the two filter resolved images is pessimistic because in reality we would have additional extra unresolved colour information used to produce the composite spectrum.
We therefore calculate the residuals between the PSF image obtained using the composite spectrum (equal to the flux-weighted sum of the true bulge and disk PSF images) and the flux-weighted sum of the bulge and disk PSF images obtained from the linear approximations to the spectra ($I_{\rm p,b}^{\rm lin}$ and $I_{\rm p,d}^{\rm lin}$ respectively), equal to $I_{\rm p,res}$.
The residuals are added to the bulge and disk PSF images obtained from the linear approximations to the spectra
at each iteration in the fit
such that the bulge PSF becomes $I_{\rm p,b}^{\rm est}=I_{\rm p,b}^{\rm lin}+I_{\rm p,res}$,
and similarly for the disk.

The bulge PSF is then convolved with the bulge image and the disk PSF with the disk image and the best-fit galaxy parameters found by minimising the $\chi^{2}$ between the sum of these images and the true PSF-convolved galaxy image.
There are eight free parameters in the fits, compared with seven free parameters in the single filter fits. The additional free parameter is the bulge amplitude in $F_{2}$. (In practice we fit for the bulge amplitude in $F_{2}$ and in the first half of $F_{1}$ to ensure that the flux in $F_{1}$ is greater than the flux in $F_{2}$). As with the single filter fits, the disk amplitude in each filter is determined from the bulge flux in the filter and the total flux (assumed known).

The multiplicative biases on the shear for the two filter configuration are shown as a function of the ratio of the
bulge-to-disk half-light radii for different galaxy redshifts in the middle panel of Fig.~\ref{fig:bias_calibration_2filter}. For comparison the biases for the single
broad-filter
imaging are shown in the left-hand panel. The right-hand panel is described in Section~\ref{sect:bias_calib}.
The central 68\% of the $r_{\rm e,b}/r_{\rm e,d}$ distribution from the catalogue ranges from approximately 0.3 to 1.6.
The figure shows that the addition of a second narrower filter reduces the biases by more than an order of magnitude.
We see that the biases for the two filter imaging are on average well within the \emph{Euclid}-like survey requirements and thus two filter imaging on board a \emph{Euclid}-like telescope would be sufficient to correct for the colour dependence of the PSF.

\begin{figure}
\center
\epsfig{file=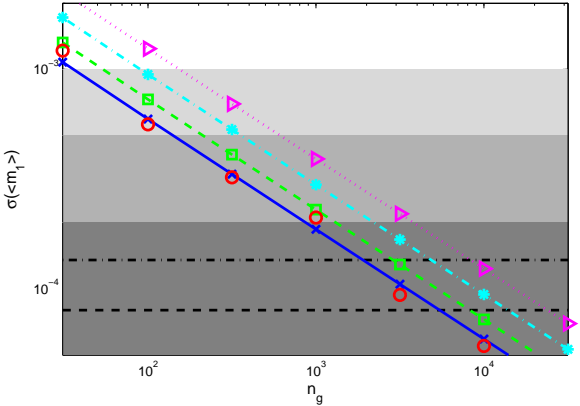,height=5.5cm,angle=0}
\caption{
Error on the mean multiplicative bias $m_{1;F_{1}}$ ($m_{2;F_{1}}$ omitted for clarity) as a function of number of galaxies in the
calibration sample
for a perfect calibration sample and
assuming the distribution in $m_{1;F_{1}}$ obtained using the catalogue is Gaussian (blue crosses solid)
and sampling from the true distribution (red circles).
Results are also shown for
a calibration sample imaged in two-filters ($F_{1}$ and $F_{2}$)
with a signal-to noise ratio of 100 (green squares dashed), 50 (cyan stars dot-dashed) and 25 (magenta triangles dotted) in the $F_{1}$ filter.
Approximate lower limit on the error on the mean bias for calibration using two filter imaging assuming a CWW-Sbc (black dashed)
and CWW-Im (black dot-dashed) spectrum for the disk.
Grey panels as in Fig.~\ref{fig:bias_filterwidth}.
}
\label{fig:ng_calibration}
\end{figure}

\section{Bias calibration}
\label{sect:bias_calib}

\begin{figure}
\center
\epsfig{file=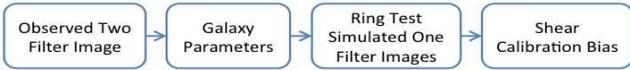,width=8.5cm,angle=0}
\caption{
Schematic of our proposal
for how to find the 1 filter bias from calibration data observed in 2 filters.
}
\label{fig:flowchart_calibration}
\end{figure}

In this section we investigate the feasibility of calibrating the biases from internal colour gradients by imaging a
subset of galaxies in more than one filter.
This subset could be obtained externally, from other existing or dedicated observations by, for example, \emph{HST}, or it could be obtained from
within the \emph{Euclid}-like survey itself, which we explore further here.
We
propose
that with real data we would follow the procedure shown in Fig.~\ref{fig:flowchart_calibration} in which the parameters of each galaxy in the
calibration sample are obtained from noisy images of the galaxy observed in two filters.
We assume that the galaxy model is known and use
the estimated parameters
to simulate the galaxy as it would be observed in a single wide filter.
Each galaxy is simulated in a ring-test to obtain the shear calibration bias.

We first estimate the number of galaxies required to calibrate the bias from single
broad-filter
imaging assuming perfect information about the galaxy properties and spectra in the calibration sample (Section~\ref{subsect:perfect_calibration}).
We then consider in Section~\ref{subsect:2filter_calibration} how well we can estimate the bias from single
broad-filter
imaging using information from noise-free images of galaxies observed in the $F_{1}$ and $F_{2}$ filters. Finally, in Section~\ref{subsect:noisy_2filter_calibration} we investigate the effects of adding noise to the two-filter images of the calibration data.

\subsection{Idealised calibration data}
\label{subsect:perfect_calibration}

Here we estimate the number of galaxies required to calibrate the bias on cosmic shear measurements from single
broad-filter
imaging. For a tomographic weak lensing analysis the galaxies will be divided up into ten or more redshift bins.
To correct the bias we need only the mean bias on the shear at each redshift,
and not the bias of each individual galaxy
(see Appendix).

Consider the optimistic case in which we have perfect information for each galaxy in a calibration sample, for example the spectrum at each point on each galaxy image. This information could be used to estimate the 1 filter shear measurement bias for each galaxy in the calibration sample. The average of these numbers could then be used to correct the shear power spectra measured from 1 filter imaging of a larger sample. Even though the information on each galaxy is perfect, this could only work if the calibration sample were a fair sample of the
Universe, and contained enough galaxies.

We use the distribution of bias values for single
broad-filter
imaging computed in Section~\ref{subsect:gal_cat} from the catalogue to
represent the full population of galaxies in a given redshift bin.
We draw $n_{\rm g}$ random samples from the distribution in $m$ to estimate the one sigma uncertainty on the mean multiplicative shear bias
for a sample of $n_{\rm g}$ galaxies. The results are shown as a function of $n_{\rm g}$ in Fig.~\ref{fig:ng_calibration}
as red circles.

We also show the results assuming the distribution in $m$ is Gaussian, so that $\sigma(\langle m \rangle )=\sigma(m)/(n_{\rm g}-1)^{0.5}$ (blue line).
The results are similar, showing that the distribution in $m$ is well represented by a Gaussian.
Therefore for a calibration sample in which the galaxy shapes and spectra are known perfectly we would need
$3\times10^3$ galaxies to calibrate the bias on the shear to a factor of 10 better than the \emph{Euclid}-like survey requirement.

\subsection{Two-filter calibration data}
\label{subsect:2filter_calibration}

We now consider how well the bias from single
broad-filter
imaging can be calibrated when the galaxies in the calibration sample are
imaged in just two filters.
In this case we will not know the bulge and disk spectra precisely, but only linear fits to the bulge and disk fluxes in each filter. There
may also be some error on the galaxy shape parameters due to the approximate spectra.

\begin{figure*}
\center
\epsfig{file=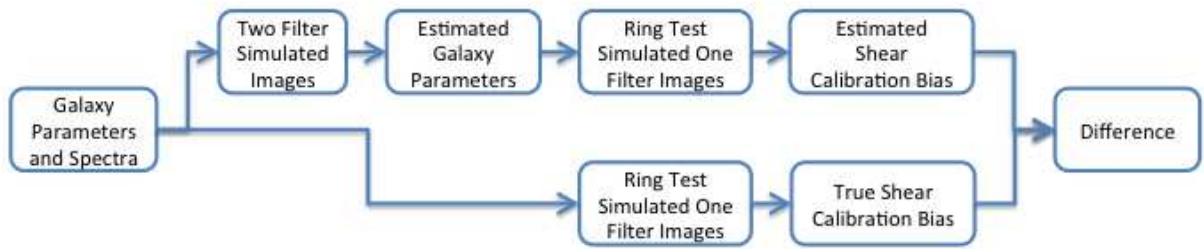,width=17cm,angle=0}
\caption{
Flowchart to explain how we test whether two filters are sufficient to do the calibration. One filter shear calibration biases from the estimated galaxy parameters are compared to those from the true galaxy parameters. The estimated galaxy parameters include estimated spectra.
}
\label{fig:flowchart_residual}
\end{figure*}

The procedure is outlined in the flow diagram in Fig.~\ref{fig:flowchart_residual}.
We first simulate the `average' galaxy from the catalogue (see Section~\ref{subsect:fid_gal_mod})
using a CWW-E spectrum for the bulge and either a CWW-Sbc or CWW-Im spectrum for the disk.
We convolve the bulge with the bulge PSF and the disk with the disk PSF (i.e. simulate the true PSF-convolved galaxy image).
The shape parameters of the galaxy and linear approximations to the bulge and disk spectra are computed using two filter imaging in $F_{1}$ and $F_{2}$ as in Section~\ref{sect:2fil}.
This is the information that we would obtain for a given galaxy in our calibration sample that could be used to estimate the bias on the shear for single
broad-filter
imaging.
We thus simulate the galaxy using this information and perform a ring-test as in Section~\ref{subsect:fil_width} to obtain an estimate for the bias on the shear when the galaxy is imaged in the $F_{1}$ filter only.

We investigate the difference between the true $m_{F_{1}}$ and $m_{F_{1}}$ measured from the calibration sample
for a range of bulge-to-disk half-light radii and galaxy redshifts.
The results are shown in the right-hand panels of Fig.~\ref{fig:bias_calibration_2filter} for
noise-free galaxy images.
The error on $m_{F_{1}}$ from calibrating in two filters depends on the ratio of the bulge-to-disk half-light radii, the disk spectrum and
the galaxy redshift.

We estimate the mean calibration error (i.e. average difference between the true and estimated $m_{F_{1}}$) over a population of galaxies.
We create a population of galaxies at a given redshift with a CWW-Sbc spectrum for the disk, and the usual CWW-E spectrum for the bulge, and use the ratio of bulge to disk sizes from the
catalogue.
We calculate the error on $m_{F_{1}}$, as shown in the right panels of Fig.~\ref{fig:bias_calibration_2filter}.
For practical purposes we restrict our calculation to the range of values on Fig.~\ref{fig:bias_calibration_2filter} and
if $r_{\rm e,b}/r_{\rm e,d}$ is less than the minimum (maximum) value on the $x$-axis of the plot then the error is set equal to the value at the minimum (maximum) $x$-value shown.
We repeat the calculation for each of the four redshift values plotted and estimate the
mean error on $m_{F_{1}}$ over redshift by weighting the contribution from galaxies at each redshift using the galaxy redshift
distribution from \citet{smailetal94}:
\(P(z)\propto z^{\alpha} \mathrm{exp}[-(1.41z/z_{\rm m})^\beta],\) where $\alpha=2$, $\beta=1.5$
and the median redshift for a \emph{Euclid}-like survey, $z_{\rm m}=0.9$.
We repeat the exercise using the CWW-Im spectrum for the disk.
The error on $m_{F_{1}}$ using the CWW-Sbc and CWW-Im spectra are shown by the lower
dashed and upper dot-dashed horizontal lines respectively in Fig.~\ref{fig:ng_calibration}.
The error on the measured bias $m_{F_{1}}$ cannot be reduced below the values shown by
these horizontal lines by increasing the number of galaxies in the calibration sample.

\subsection{Noisy two-filter calibration data}
\label{subsect:noisy_2filter_calibration}

We repeat the calculations of the previous section but adding noise to the two-filter simulated images.
The noise causes a spread of estimated galaxy parameters and thus increases the width of the
distribution
of $m$ values and could induce an extra bias.

A broader distribution of $m$ values means that more galaxies will be needed in the calibration sample to measure the mean $m_{F_{1}}$ to the required accuracy.
We add the variance of the $m$ distribution obtained from the
catalogue
with the variance of the $m$ distribution for the `average' galaxy from the catalogue whose parameters were obtained from two-filter imaging of noisy images of the galaxy.
The  number of galaxies required in the calibration sample for different signal-to-noise values is shown in Fig.~\ref{fig:ng_calibration}.
The signal-to-noise is the value measured in the $F_{1}$ filter. We assume the same exposure time in the $F_{1}$ and $F_{2}$ filters and so the signal-to-noise in $F_{2}$ is less than the values quoted.
Note for illustration we assume that for a given galaxy the broadening of $m_{F_{1}}$ due to noise is independent of the galaxy shape parameters and colour gradient and equal to the broadening for the `average' galaxy.

We check that the mean bias for single filter imaging in $F_{1}$ estimated from two filter imaging of noisy calibration data
is not wrong by more than the requirement on the bias.
We find that the difference between the true bias from imaging in a single wide filter and the mean estimated bias over
many noise realisations is at least a factor of two less than the \emph{Euclid} requirement down to a signal-to-noise of 25.
We note
that there is no indication that the bias difference increases as the SNR decreases from 100 to 25.

\section{Discussion}
\label{sect:discussion}

To capitalise on gravitational lensing as a cosmological probe it can be attractive to use wide-filter imaging to gather as much light as possible and constrain
cosmological parameters, including the dark energy equation of state parameters, to high precision. Such
wide-filter
experiments may however cause an
unacceptable loss in accuracy. In this paper we consider the shear measurement biases arising from single filter imaging due to the wavelength-dependence of the PSF and the existence of colour gradients across galaxy images.

The PSF is modelled as a Gaussian at each wavelength with ellipticity 0.05.
We have assumed a particular wavelength dependence of the PSF size.
In practice different telescope designs will have different
dependencies depending mainly on the mirror size, the amount of jitter and the amount of charge diffusion.  As can be seen from \citet{cypriano10}, the
optics dominate the PSF wavelength dependence and the biggest likely change would be the relative contribution of the other two (relatively
wavelength independent) terms. A decrease in the jitter and/or amount of charge diffusion would increase the wavelength dependence of the PSF. In the
worst possible case of zero charge diffusion and no jitter, the slope of the PSF size as a function of wavelength is approximately doubled, which we
would naively expect to double the shear measurement biases. In practice the change would be smaller, and if the diffusion or jitter were larger
than assumed then the biases would be smaller.
We have assumed that the PSF ellipticity is independent of wavelength, but in practice additional components such as beam splitters could
introduce wavelength dependent ellipticity.
We assume that the ellipticity, orientation and wavelength-dependence of the size of the PSF are all known.

We consider simple bulge plus disk galaxies modelled using two co-centered, co-elliptical S\'{e}rsic profiles, each with a different spectrum.
The bulge is modelled as a de Vaucouleurs profile and the disk as an exponential.
We assume the galaxy model is known and fit for the centre, ellipticity and orientation of the galaxy,
the sizes of the bulge and disk components and the bulge-to-total flux ratio. We also assume the composite spectrum emitted by the galaxy is
known from ground-based observations and do not take into account errors on the shape of the spectrum or total flux emitted.
In practice the different passbands and apertures used from ground and space will make it difficult to obtain the true total flux which will make
the estimated biases noisier and potentially biased.

The results will depend on the complexity of the galaxy model.
For example it is possible that non-elliptical isophotes of galaxies can produce non-negligible additive shear calibration biases (Hirata \& Bernstein priv. comm.).
However an investigation using more complex models than the
elliptical isophote model considered here is beyond the scope of this work.
In particular we note that extending the calculation to more complex galaxy models is non-trivial because current shape measurement methods
do not meet the accuracy requirements for a \emph{Euclid}-like survey for realistic signal-to-noise galaxies with radially-varying ellipticity isophotes
\citep[][]{great08resultsmnras}; though considerable progress has been made on
understanding the limitations of existing methods \citep[][]{voigt&bridle10,melchior2010} and on the development of new methods
\citep[][]{bernstein2010,great10handbook}.

We first measure the bias on the shear if we ignore colour gradients and image galaxies in a single filter.
In the simulations of the `true' galaxy images
we model the bulge spectrum using a CWW-E spectrum and the disk using either a CWW-Sbc or CWW-Im spectrum.
We do not take into account evolution of these spectra with redshift as this is much smaller than the difference between the CWW-Sbc and CWW-Im spectra and we show results for both.
In the fits to the simulated images
we assume the spectrum at each point in the galaxy image is equal to the composite spectrum from the bulge plus disk emission.

We measure the dependence of the bias on the width of the filter for example galaxy parameters. The bias drops by a factor of
approximately 5 when the filter width is halved.
The bias depends strongly on the galaxy shape parameters as well as on the colour gradient.
For a `typical' galaxy and a filter width of 350 nm the bias is a factor of 3 below the \emph{Euclid}-like survey
systematic equals random error
requirement
of $10^{-3}$,
but the bias can be more than a factor of 10 above the requirement for some galaxy properties considered.

We show that, to first order, tomographic cosmic shear two point statistics depend on the mean shear bias over the
galaxy population at a given redshift.
We compute the mean bias for single
broad-filter
imaging, averaging over a catalogue of galaxies
from a published study of \emph{HST} observations
of galaxies in the Keck LRIS DEEP1 survey
\citep{simard02}.
The catalogue is used to
represent the range of galaxy colour gradients, shape parameter ratios $r_{\rm e,b}/r_{\rm e,d}$ and
bulge-to-total flux ratios in the Universe.
We caution, however, that the catalogue has a complicated selection function (see Section~\ref{subsect:gal_cat}),
and that this must be taken into account when drawing conclusions for future cosmic shear surveys.
We attempt to investigate the sensitivity of the results to the catalogue by
plotting the bias dependence on different galaxy parameters, and by re-computing the mean bias
for different colour cuts on the catalogue.

We find that the
mean
bias
over the full catalogue
is approximately a factor of
3 below the \emph{Euclid}-like survey requirement.
(The mean bias over the catalogue is similar to the bias for the fiducial galaxy because the parameters of the fiducial galaxy are chosen to represent
the `average' galaxy from the catalogue).
However,
to ensure the total bias arising from all shear measurement systematics is below the
\emph{Euclid}-like requirement, the mean bias from the colour dependence of the PSF
should be of order a factor of 10
below the
systematic equals random error
requirement.
We thus conclude in Section~\ref{sect:single_filter} that single broad-filter imaging in a future \emph{Euclid}-like survey
may not allow the
accuracy requirements on the bias to be met,
depending on how accurately the catalogue from \citet{simard02} represents the cosmic shear survey being considered.

We assume that the mean bias obtained using the full catalogue is optimistic because we find it may be up to a factor of 2 larger
than $10^{-3}$ if we modify the galaxy redshift distribution by making
colour cuts on the catalogue.
However, it is possible that the mean bias computed using galaxy parameter distributions from the catalogue
is larger than the true mean bias in a \emph{Euclid}-like survey.
This is because the photometric and structural parameters obtained in the \citet{simard02} study will not precisely represent
the underlying distribution of colours and shapes of galaxies in the DEEP1 survey as a result of fitting a simplified model to noisy galaxies
\citep[see][]{haussleretal07}.
It is likely that this will result in extreme bulge-to-total flux ratios and colours, increasing the mean bias.

The bias from single broad-filter imaging can be reduced to a negligible level by narrowing the filter,
however the survey time or efficiency would have to increase to maintain the same cosmological constraining power.
We also consider two
other
possible options: (i) multiple-filter imaging of the whole survey and (ii) calibrating the bias
from single broad-filter imaging
on a sub-set of galaxy images observed in more than one filter. We first investigate how much the biases may be reduced by imaging in two filters. We consider an additional narrower filter with width equal to half the width of the
wide
filter and fully contained within the wide filter. The colour information available from a two-filter configuration allows us to determine linear approximations to both the bulge and disk spectra across the
wide
filter.
Using the bulge and disk PSF models calculated from these
linear
spectra,
and including the information from the composite spectrum,
the biases are reduced by an order of magnitude below the biases from single
broad-filter
imaging. Thus,
for the simple bulge plus disk galaxy model considered,
only one narrow filter is needed in addition to the wide filter to remove the bias from the PSF wavelength dependence to the required accuracy.
We note that if the galaxy has more than two components, each with a different spectrum, then
two-filter imaging is sufficient to estimate linear approximations to the spectra of each of the components, provided the components are resolved.

We also consider the possibility of calibrating the bias from single
broad-filter
imaging on a sub-set of galaxy images.
For a tomographic cosmic shear analysis
we need the mean multiplicative bias for each redshift bin to be measured to an accuracy of $10^{-4}$ (i.e. a factor of ten below the \emph{Euclid}
requirement).
We note that we expect the mean bias to be redshift-dependent because galaxy morphologies and spectra evolve with redshift.
We first compute the number of
galaxies needed in the calibration sample
to calibrate a \emph{Euclid}-like survey
assuming perfect knowledge of the galaxy shape parameters and spectra.
Using the distribution of galaxy shapes and colour gradients in the catalogue as a representation of the population of galaxies in the Universe
we find that we need approximately $3\times10^{3}$ galaxies per redshift bin.

We then investigate whether the information available from noisy two-filter imaging is sufficient to calibrate the bias from single
broad-filter
imaging.
We assume a constant exposure time in the two filters so that the signal-to-noise in the narrower filter is lower than in the wide filter.
We find that for a signal-to-noise in the wide filter equal to 25 the number of galaxies required to calibrate the bias in each tomographic redshift bin is of order $10^{4}$.

We caution, however, that the calibration strategy assumes the galaxy model is known
and further investigation is required to determine whether the procedure will work on low signal-to-noise images of
real galaxies with complex morphologies.
We check that the error on the calibration bias measured from two-filter imaging of noisy galaxies is sufficiently small down to a signal-to-noise of 25.
We do not check lower signal-to-noise values due to computational limitations.
We note that in practice it may be possible to stack images with similar properties, e.g. colours,
within each redshift bin
to reduce the noise.

We have used a simple galaxy model in our proposed calibration strategy because this may capture the main colour gradient effect that causes the shear measurement bias. However this may be insufficiently flexible in practice, depending on the complexity of galaxies in the Universe and any additional shear measurement bias this causes. For this simple galaxy model we find that the shear measurement bias is relatively insensitive to the true galaxy ellipticity (lower left panel of Fig~\ref{fig:galpar}), and therefore galaxies from a calibration sample could be imaged with lower PSF quality than for the main cosmic shear sample. Qualitatively this is because the calibration sample provides information on galaxy colour gradients which has a less stringent image quality requirement than the very stringent requirement for cosmic shear.

We note that the bias value we obtain from the galaxy catalogue cannot be taken to be the true bias for a \emph{Euclid}-like survey because it contains an insufficient number of galaxies and the selection criteria are not matched. The sample contains 632 galaxies over a range of redshifts whereas we conclude that a
calibration
sample of around $10^{4}$
galaxies
per tomographic redshift bin
is required.
Our conclusions on the number of galaxies needed in a calibration sample should be more robust though, because the catalogue gives us information about the range of galaxy types in the Universe.

We note that around $r_{\rm b}=r_{\rm d}$ the multiplicative bias changes sign.
We expect the bias to be zero at $r_{\rm b}=r_{\rm d}$ for the simple case where the spectral index of the bulge and
disk are the same, because the radial intensity profile would be identical for each component and thus the true PSF-convolved galaxy image equal to the
true galaxy image convolved with the PSF for the composite spectrum.
We find that when the bulge and disk spectral indices are different the bias is negligible for $r_{\rm b}/r_{\rm d}$ close to 1.
We note that the average of the Simard catalogue
occurs near to this point ($r_{\rm b}/r_{\rm d}=1.1$) and
therefore this contributes significantly to the small bias obtained due to the cancellation of large biases of either sign. Again, if the Simard
catalogue is atypical then this could lead to a larger bias from realistic data.

\section*{Acknowledgments}
We are grateful to Lance Miller, Gary Bernstein, Catherine Heymans, Henk Hoekstra, Stefano Andreon, Eduardo Cypriano, Marcella Carollo, Ignacio Ferreras, Ewan Cameron, Alexandre Refregier, Michael Seiffert, Kevin Bundy, Stephane Paulin-Henrikkson, Anais Rassat, Donnacha Kirk, Ole Host, Filipe Abdalla, Ofer Lahav, Chris Hirata and Sam Thompson for helpful discussions.
LMV acknowledges support from the STFC.
SLB thanks the Royal Society for support in the form of a University
Research Fellowship and the European Research Council for support in the form of a Starting Grant.
TDK is support by a RAS 2010 Fellowship.
RJM thanks STFC for support in the form of an Advanced Fellowship.
JR was supported by the Jet Propulsion Laboratory, operated by the California Institute of Technology under a contract for NASA.
TS acknowledges support from NSF through grant AST-0444059-001, and
the Smithsonian Astrophysics Observatory through grant GO0-11147A.

\bsp
\bibliographystyle{mn2e}


\section*{Appendix: Impact on shear power spectra of density dependent shear calibration}

Galaxies with a particular colour gradient will tend to be clustered, for example clusters of galaxies contain a lot of ellipticals which tend to have a small colour gradient, and the converse will be true in voids. In this Appendix we check how this will affect the tomographic lensing power spectrum.

This calculation is related to that by
\citet{guzik&bernstein2005}
who calculate the effect of spatially varying multiplicative shear calibration biases which are uncorrelated with the shear. Here we assume that the multiplicative bias is correlated with the density perturbations and therefore to the shear. There are some parallels between this Appendix and calculations of the effect of galaxy number density weighted shear measurements
\citep[e.g. see][]{hiratas04}
or source-lens clustering
\citep{Bernardeau1998,Hamanaetal2002,Forero-Romeroetal2007},
and the shear-intrinsic alignment or GI effect
\citep{hiratas04}.
However, the former calculations do not consider shear calibration bias.

In this derivation we assume that galaxies have a stochastic multiplicative shear measurement bias $m$ so that the observed shear $d$ is related to the true shear $\gamma$ by
\begin{equation}
d = (1+m)\gamma
\end{equation}
where $m$ is drawn from a probability distribution whose shape depends on the spatial location
$\Pr(m|\btheta,z)$.
Here we assume that the whole probability distribution is stretched or squeezed uniformly according to the local matter density
\begin{equation}
\Pr(m|\delta) = f((1+b_{m}\delta)m)
\end{equation}
where $b_{m}$ quantifies the sensitivity of the multiplicative bias to the mass distribution. We would expect $b_{m}$ to be negative to obtain the expected reduced multiplicative bias in a local overdensity.  Here we choose to assume $b_{m}$ is independent of scale but is an arbitrary function of redshift $b_{m}(z)$, which would depend on galaxy evolution.
It would be possible to generalise this derivation for density independent shear calibration biases discussed in
\citet{guzik&bernstein2005}
by adding in terms to the above equations, but we choose to focus the discussion here on the new effect.

For first order calculations of the shear two point statistics the relevant property of this distribution will be the mean,
$\bar{m}$, which can be written
\begin{equation}
\bar{m}(\btheta,z) = (1+b_{m}(z)\delta(\btheta,z))\bar{\bar{m}}(z)
\end{equation}
where $\bar{\bar{m}}$ is the mean multiplicative shear calibration bias averaged over the sky at a given redshift, since $\langle\delta\rangle=0$.
The description of the probability distribution could therefore be more general than described in the paragraph above.
Note that to achieve the usual effect of decreased biases at high densities, $b_m$ needs to be negative.
At very high densities this equation predicts very negative multiplicative shear biases, which would fit with very blue cores in galaxies in the centres of clusters. This is not particularly physical and therefore the model should be improved by using a more sophisticated function of $\delta$ which could affect the cancellations in the final result.
For example some inverse function of $(1+\delta)$ could have a more realistic effect at high densities.

The main cosmic shear statistic of interest for future surveys is the tomographic shear power spectrum, which may be defined in terms of the true shear $\gamma$
\begin{equation}
C_{\ell i j}^{GG} \equiv \langle \langle \tilde{\gamma}_{\mathbf{\ell} i} \tilde{\gamma}_{\mathbf{\ell} j}^{*} \rangle_{\phi \mathbf{\ell}} \rangle_{\rm{r}}
\end{equation}
where the inner angle bracket denotes an average over direction of the vector $\mathbf{\ell}$ and the outer angle bracket over realisations. We use $G$ to denote true gravitational shear. Here $i$ and $j$ denote the index of the tomographic redshift bins used for the correlation. Tilde denotes the usual Fourier transform in the plane of the sky
\begin{equation}
\tilde{x}_{\mathbf{\ell}} \equiv  \int x(\btheta)  e^{i\mathbf{\ell}. \btheta}  .
\end{equation}

The shear $\gamma$ is related to the mass density $\delta$ by the usual equations
\begin{equation}
\gamma(\btheta,z) = D(\btheta) * \kappa(\btheta,z)
\end{equation}
where
\begin{equation}
\kappa(\btheta,z) = \int dz' q(z,z') \delta(\btheta,z')
\end{equation}
using the Limber approximation and writing the three dimensional mass density per unit comoving volume in terms of position on the sky and redshift.
$q$ is the usual lens weight function but the only relevant property of $q$ for this paper is that $q(z,z')$ is zero when $z=z'$.

We assume that tomographic redshift bins are predetermined according to some observable property which is independent of the density field so that the number density of galaxies in redshift bin $i$ as a function of position on the sky and redshift $n_i(\btheta,z)$ may be separated out as
\begin{equation}
n_i(\btheta,z) = n_i(z) (1+\delta_g(\btheta,z))
\end{equation}
where $n_i(z)$ is the mean number density of galaxies at redshift $z$ in redshift bin $i$ averaged over a patch of sky large enough that the mean galaxy density fluctuation $\delta_g(\btheta,z)$ is zero, normalised such that
\begin{equation}
\int n_i(z) dz = 1 .
\end{equation}

We can only calculate statistics of the observed shear $d$ and we here assume this is done by producing a map of $d$ on the sky for each redshift bin. We assume this is done by averaging observed shears in pixels on the sky, therefore this map can be written in the limit of infinite source density as
\begin{equation}
d_i(\btheta) = \int dz \int dm \,\, n_i(\btheta,z) \Pr(m|\btheta,z) d(m,\btheta,z)
\end{equation}
which can be written to first order in the small quantities as
\begin{equation}
d_i(\btheta) = \gamma_i(\btheta) + \gamma_{g i}(\btheta) + \gamma_{m i} (\btheta)
\end{equation}
where we have defined
\begin{eqnarray}
\gamma_i (\btheta) &=& \int dz \,\, n_i(z) \gamma(\btheta,z) \\
\gamma_{g i}(\btheta) &=& \int dz \,\,n_i(z) \delta_g(\btheta,z) \gamma(\btheta,z) \\
\gamma_{m}(\btheta) &=&
\int dz \,\,n_i(z) \int dm \Pr(m|\btheta,z) m dm \gamma(\btheta,z) \\
&=&
\int dz \,\,n_i(z) \bar{m}(\btheta,z) \gamma(\btheta,z) .
\end{eqnarray}

We will measure the power spectrum of the observed shear $d$
\begin{equation}
C_{\ell i j}^{dd} \equiv \langle \langle \tilde{d}_{\mathbf{\ell} i} \tilde{d}_{\mathbf{\ell} j}^{*} \rangle_{\phi \mathbf{\ell}} \rangle_{\rm{r}} .
\end{equation}
which can be written to first order in the perturbative quantities as
\begin{equation}
C_{\ell i j}^{dd}  = C_{\ell i j}^{GG}  + C_{\ell i j}^{Gg} + C_{\ell i j}^{Gm} + C_{\ell i j}^{gG} + C_{\ell i j}^{mG}
\end{equation}
where we have defined
\begin{eqnarray}
C_{\ell i j}^{Gg} &=&   \equiv \langle \langle
\tilde{\gamma}_{\mathbf{\ell} i} \tilde{\gamma}_{g \mathbf{\ell} j}^{*} \rangle_{\phi \mathbf{\ell}} \rangle_{\rm{r}} \\
C_{\ell i j}^{Gm} &=&   \equiv \langle \langle
\tilde{\gamma}_{\mathbf{\ell} i} \tilde{\gamma}_{m \mathbf{\ell} j}^{*} \rangle_{\phi \mathbf{\ell}} \rangle_{\rm{r}}
\end{eqnarray}
where $C_{\ell i j}^{Gg}$ is the same as the term written with the same notation in
\citet{joachimib10}
but $C_{\ell i j}^{Gm}$ is different because in this paper we use $m$ to denote the multiplicative shear calibration bias whereas in that work $m$ was used to denote magnification which we ignore here.

The first term above is the usual shear term and the only one usually considered, the second takes into account the fact that we measure the galaxy number density weighted shear (which contributes to source-lens clustering) and the third term takes into account the possible dependence of the measured shear on a multiplicative shear measurement calibration bias. We write down the source-lens clustering term here because the effect  may at first sight seem correlated with the effect discussed in this paper, however from the above equations it can be seen that to the level of approximation considered here the terms are independent, and we thus discuss it no further.
Note that the final two terms can be non-zero for overlapping redshift bins.

The cross shear-multiplicative bias term may be further expanded as
\begin{equation}
C_{\mathbf{\ell} i j}^{Gm} = C_{\mathbf{\ell} ij}^{G \bar{\bar{m}}} + C_{\mathbf{\ell} ij}^{Gmb}
\end{equation}
where we have defined for convenience
\begin{eqnarray}
\gamma_{\bar{\bar{m}} i}(\btheta) &=& \int dz \,\,n_i(z) \bar{\bar{m}}(z) \gamma(\btheta,z) \\
\gamma_{mb i}(\btheta) &=& \int dz \,\,n_i(z) \bar{\bar{m}}(z) b_m(z) \delta(\btheta,z) \gamma(\btheta,z) \\
C_{\mathbf{\ell} ij}^{G \bar{\bar{m}}}
&=&   \equiv \langle \langle
\tilde{\gamma}_{\mathbf{\ell} i} \tilde{\gamma}_{\bar{\bar{m}} \mathbf{\ell} j}^{*} \rangle_{\phi \mathbf{\ell}} \rangle_{\rm{r}} \\
C_{\mathbf{\ell} ij}^{Gmb}
&=&   \equiv \langle \langle
\tilde{\gamma}_{\mathbf{\ell} i} \tilde{\gamma}_{mb \mathbf{\ell} j}^{*} \rangle_{\phi \mathbf{\ell}} \rangle_{\rm{r}}.
\end{eqnarray}

We now focus on the second term above and substitute in the relation between shear and mass density. We find
\begin{equation}
\tilde{\gamma}_{\bar{\bar{m}} \mathbf{\ell} i}
= \int dz \, n_i(z) \bar{\bar{m}}(z) b_m(z) \int dz' \, q(z',z)  A_{\mathbf{\ell}}(z,z')
\end{equation}
where we have defined
\begin{equation}
A_{\mathbf{\ell}}(z,z') = \tilde{\delta}_{\mathbf{\ell}}(z) * \left(\tilde{D}_{\mathbf{\ell}} \tilde{\delta}_{\mathbf{\ell}}(z')\right).
\end{equation}
Therefore
\begin{equation}
C_{\mathbf{\ell} ij}^{Gmb} = \int dz \, g_i(z) \int dz' \, n_j(z') \bar{\bar{m}}(z') b_m(z') \int dz'' \, q(z'',z') B_{D*\mathbf{\ell}} (z,z',z'')
\end{equation}
where we defined
\begin{equation}
B_{D*\mathbf{\ell}} (z,z',z'')
  \equiv \langle \langle
  \tilde{D}_{\mathbf{\ell}}
  \tilde{\delta}_{\mathbf{\ell}}(z)
  \left(
    \tilde{\delta}_{\mathbf{\ell}}(z'')
    *
    \left(
      \tilde{D}_{\mathbf{\ell}}
      \tilde{\delta}_{\mathbf{\ell}}(z')
    \right)
  \right)^*
  \rangle_{\phi \mathbf{\ell}} \rangle_{\rm{r}}.
\end{equation}

If we assume that mass at one redshift is uncorrelated with mass at another redshift then $B_{D*\mathbf{\ell}} (z,z',z''))$ is zero unless $z=z'=z''$. Taken together with the fact that $q(z,z')=0$ when $z=z'$ it would follow that $C_{\mathbf{\ell} ij}^{Gmb} =0$ at all redshifts. In any case this term depends on the three point function of the mass distribution and is expected to be even smaller than the source-lens clustering term due to the additional factor of $\bar{\bar{m}}$.

The remaining term $C_{\mathbf{\ell} ij}^{G \bar{\bar{m}}}$ is much simpler and can be written
\begin{equation}
C_{\mathbf{\ell} ij}^{G \bar{\bar{m}}} = \int dz \, g_i(z) g_{mj}(z) P(k=\ell/\chi,z)
\end{equation}
where we have defined
\begin{equation}
g_{m i}(z) \equiv \int dz' n_i(z) \bar{\bar{m}}(z') q(z,z')  ,
\end{equation}
$\chi$ is the comoving distance to redshift z and $k$ is the comoving wavenumber.
In the special case where $\bar{\bar{m}}$ is independent of redshift then
\begin{equation}
C_{\mathbf{\ell} ij}^{G \bar{\bar{m}}} = \bar{\bar{m}} C_{\mathbf{\ell} ij}^{GG}
\end{equation}
as expected for the simple case where $d = (1+ \bar{\bar{m}})\gamma$ (note the additional term
$C_{\mathbf{\ell} ij}^{\bar{\bar{m}} G}$ which makes up the factor of 2).

Therefore we have found that the cosmic shear power spectrum is modified to first order by a term proportional to the average of the multiplicative shear calibration bias. In detail there is an additional power spectrum which is identical to the usual lensing power spectrum but in which the lensing weight function is weighted by the mean multiplicative shear calibration bias m at each redshift, spatially averaged over the sky. This is why we require for calibration of this effect the mean multiplicative shear calibration factor at each redshift. This should be spatially averaged rather than galaxy number density averaged.
The number of redshift bins in which $\bar{\bar{m}}$ should be calculated depends on how smoothly it changes with redshift.

\begin{equation}
\end{equation}

\label{lastpage}

\end{document}